\documentclass[prd,showpacs,preprintnumbers,amsmath,amssymb,floatfix]{revtex4-2}

\usepackage{graphicx}
\usepackage{dcolumn}
\usepackage{bm}
\usepackage{amssymb}
\usepackage{epsfig}
\usepackage{color}

\newcommand{\ba}{\begin{eqnarray}}
\newcommand{\ea}{\end{eqnarray}}
\newcommand{\be}{\begin{equation}}
\newcommand{\ee}{\end{equation}}
\newcommand{\bdisplay}{\begin{displaymath}}
\newcommand{\edisplay}{\end{displaymath}}

\newcommand{\eq}[1]{Eq.\,(\ref{#1})}
\newcommand{\fig}[1]{Fig.\,\ref{#1}}

\makeatletter
\def\eqnarray{\stepcounter{equation}\let\@currentlabel=\theequation
\global\@eqnswtrue
\tabskip\@centering\let\\=\@eqncr
$$\halign to \displaywidth\bgroup\hfil\global\@eqcnt\z@
  $\displaystyle\tabskip\z@{##}$&\global\@eqcnt\@ne
  \hfil$\displaystyle{{}##{}}$\hfil
  &\global\@eqcnt\tw@ $\displaystyle{##}$\hfil
  \tabskip\@centering&\llap{##}\tabskip\z@\cr}

\def\endeqnarray{\@@eqncr\egroup
      \global\advance\c@equation\m@ne$$\global\@ignoretrue}

\def\@yeqncr{\@ifnextchar [{\@xeqncr}{\@xeqncr[5pt]}}
\makeatother

\begin{document}

\title{Some applications of the eikonal model with Coulomb and curvature corrections in $pp$ and $\bar{p}p$ scattering}
\author{Phuoc Ha}
\email{pdha@towson.edu}
\affiliation{Department of Physics, Astronomy and Geosciences, Towson University, Towson, MD 21252}

\begin{abstract}
Using a simple eikonal approach to the treatment of Coulomb-nuclear interference and form-factors effects and taking into account the curvature effects in high-energy $pp$ and $\bar{p}p$ scattering, we determine the basic parameters $B$, $\rho$, and $\sigma_{\rm tot}$ from fits to experiment at $W=\sqrt s=$ 53 GeV, 62.3 GeV,  8 TeV, and 13 TeV. We then investigate the differential cross sections in the dip region for $pp$ and $\bar{p}p$  elastic scattering at $W=$ 53 GeV and 1.96 TeV. We find that the results of the basic parameters calculated using the simple eikonal approach agree well with the values determined in other analyses. We also find that Coulomb effects are significant in the dip region at 53 GeV and 1.96 TeV, and must be taken into account in searches for odderon effects through cross section differences in that energy region. 
\end{abstract}
%
%
\maketitle


\section{Introduction \label{sec:introduction}}

In a recent paper \cite{DH2020}, we presented an analysis of the Coulomb and form-factor effects in $pp$ scattering based on an eikonal model for the spin-averaged $pp$ scattering amplitude. We note that the study of the Coulomb-nuclear interference and its use to extract $\rho$ has a long history \cite[and many others]{BetheCoulomb,WestYennie,Cahn,Islam,KL-Coulomb,Buttimore,KopeliovichTarasov} and that the general nucleon-nucleon scattering theory is reviewed in \cite{Blockrev, BlockCahn} and the status leading up to the LHC is reviewed in \cite{PanSri}.

Our eikonal approach  \cite{eikonal2015} was based on a realistic model which fitted the $pp$ and $\bar{p}p$ data from 4.5 GeV to cosmic ray energies, and was consistent with the phase constraints imposed by analyticity \cite{Blockrev}.  The model allowed us to calculate the Coulomb and form-factor effects in the scattering without significant approximation at any value of $q^2$ for which it held, extending beyond the first diffraction minimum in the differential cross section. 

In that approach, the spin-averaged $pp$ scattering amplitude was given at small $q^2$ by
\be
\label{final_amp}
 f(s,q^2) = -\frac{2\eta}{q^2}F^2(q^2) + e^{i\Phi_{tot}(s,q^2)}f_N(s,q^2).
 \ee
The advantage of this form of the amplitude was that the Coulomb term was real, making it clear that the Coulomb-nuclear interference depended only on the real part of the second term, that was, on the real part of $f_N(s,q^2)$ with a (small) admixture of the imaginary part dependent on the phase $\Phi_{tot}(s,q^2)$. The latter was essentially model- independent for any eikonal model consistent with the the measured $pp$ and $\bar{p}p$ total cross sections, the forward slope parameters $B=-d\log\left(d\sigma/dq^2\right)/dq^2$, and the diffractive structure at larger $q^2$. We found that $\Phi_{tot}(s,q^2)$ was small and easily parametrized in the small-$q^2$ region and allowed the simple extraction of $\rho$ from the data.

Here, we modify the approach discussed  in \cite{DH2020} to isolate the purely nuclear scattering amplitude, with the mixed Coulomb-nuclear effects contained in a small correction term. We use the result to analyze a model used in recent fits to Coulomb-nuclear interference at high energies  \cite{ Nagy1979, Breakstone1985, ISR1985} and very high energies \cite{TOTEM2016,TOTEM2019}. Since the Coulomb and form-factor corrections are effectively model-independent, we can proceed to a simpler construction used in various experimental analyses in which the purely nuclear part of the differential cross section is approximated as
\be
\label{curvature_exp}
\frac{d\sigma}{dq^2}(s,q^2) \approx Ae^{-Bq^2+Cq^4-Dq^6+\cdots},\quad 0\leq q^2\ll 1.
\ee
Here $B$ is the usual slope parameter and the parameters $C,\,D\,\cdots$ which introduce curvature in $d\sigma/dq^2$ are calculated by using our eikonal model fitted to the high energy $pp$ and $\bar{p}p$ data  \cite{curvature_fit, bdhh-curvature}. This reduces the number of free parameters by two relative to those used in other analyses of this type  \cite{TOTEM2016,TOTEM2019, ATLAS2022}.

Using this approach, we evaluate the basic parameters $B$, $\rho$, and $\sigma_{\rm tot}$ at $\sqrt s =$ 53 GeV, 62.3 GeV,  8 TeV, and 13 TeV and find that the results of the basic parameters calculated using our simple eikonal approach agree well with the values determined in other analyses.  We also study the differential cross sections in the dip region for $pp$ and $\bar{p}p$  elastic scattering at $W=$ 53 GeV and 1.96 TeV. We find that Coulomb effects are significant there and must be taken into account in attempts to detect odderon effects from differences in the $pp$ and $\bar{p}p$ cross sections as studied by the D0 and TOTEM collaborations \cite{ D0TOTEM2020}.


\section{ Simple Eikonal approach for Coulomb-nuclear interference effects \label{sec::analysis}}


In the absence of significant spin effects, the spin-averaged differential cross section for $pp$ and $\bar{p}p$ scattering can be written in terms of a single spin-independent amplitude
\be
\label{f^tot}
f(s,q^2) = i\int_0^\infty db\,b\left(1-e^{2i\left(\delta_c^{tot}(b,s)+\delta_N(b,s)\right)}\right)J_0(qb),
\ee
Here $q^2=-t$ is the square of the invariant momentum transfer, $b$ is the impact parameter, $\delta_c^{tot}(b,s)$ is the full Coulomb phase shift including the effects of the finite charge structure of the proton,   $\delta_N(b,s)$ is the nuclear phase shift, and
\be
\label{delta_c^tot}
\delta_c^{tot}(b,s) = \delta_c(b,s)+\delta_c^{FF}(b,s),
\ee
where $\delta_c$ gives the phase shift for a  pure Coulomb interaction, and $\delta_c^{FF}$ accounts for the effects of the form factors at large momentum transfers or short distances. 

 \eq{f^tot} can be rearranged in the form
\be
f(s,q^2) = f_c(s,q^2) + f_c^{FF} (s,q^2)+f_{N,c}(s,q^2),
\label{C+N_amp}
\ee
where
\ba
f_c(s,q^2) &=& i\int_0^\infty db\,b\left(1-e^{2i\delta_c(s,b)}\right)J_0(qb)  \\ 
f_c^{FF} (s,q^2)&=& i\int_0^\infty db\, b\, e^{2i\delta_c(b,s)}\left(1-e^{2i\delta_c^{FF}(b,s)}\right)J_0(qb)  \\
f_{N,c}(s,q^2)&=& i\int_0^\infty db\,b\,e^{2i\delta_c(b,s)+2i\delta^{FF}_c(b,s)}\left(1-e^{2i\delta_N(b,s)}\right)J_0(qb).
\ea
Here $f_c(s,q^2)$ is the  Coulomb amplitude  without form factors,  $f_c^{FF}$ accounts for the effects of the form factors on the Coulomb scattering, and $f_{N,c}$ includes the effects of the nuclear scattering as modified by the Coulomb and form factor effects.

The pure nuclear amplitude $f_N(s,q^2)$ is just
\be
\label{f_N}
f_N(s,q^2) = i\int_0^\infty db\,b\left(1-e^{2i\delta_N(b,s)}\right)J_o(qb).
\ee
We have studied this in detail in an eikonal model fitted to data on  $\sigma_{\rm tot}$, $B$,  and $\rho$ for $pp$ and $\bar{p}p$ scattering from 4.5 GeV to cosmic ray energies  \cite{eikonal2015}.

Before going further with Eq.(\ref{C+N_amp}), for $pp$ scattering, we can divide out the common Coulomb phase $(4p^2/q^2)^{i \eta}$ (or phase $(4p^2/q^2)^{-i \eta}$ for $\bar{p}p$ scattering)  from all terms; $f_c$ is then real.

 Using the standard proton charge form factor
\footnote{Magnetic-moment scattering does not contribute to the spin-independent part of the scattering amplitude. Its contribution to the spin-dependent amplitude is also suppressed  by an angular factor proportional to $\sqrt{q^2}$ at small $q^2$. The factor $1/q^2$ from the photon propagator therefore partially cancels and the amplitude is suppressed relative to the charge-scattering term for $q^2\rightarrow 0$.  The magnetic form factors $F_M^2(q^2)$  further suppress the scattering at large $q$. As a result, the magnetic terms  do not contribute significantly to the scattering in the region of interest here. See a detail discussion in \cite{DH2020} }
\be
\label{form_factors}
F_Q(q^2) =\frac{\mu^4}{(q^2+\mu^2)^2}
\ee
 with $\mu^2=0.71$ GeV$^2$, 
 for $pp$ scattering, we find that for the form-factor corrections to the Coulomb amplitude
 \be
 \label{CoulFF}
 f_c(s,q^2)+f_c^{FF}(s,q^2) = -\frac{2\eta}{q^2}\left[1 -\left(\frac{q^2}{q^2+\mu^2}\right)^{i\eta}  + \left(\frac{q^2}{q^2+\mu^2}\right)^{i\eta}\frac{\mu^8}{(q^2+\mu^2)^4}\right]
 \ee
up to negligible real contributions of order $\eta^2$.
For $\eta/F_Q^2(q^2)\ll 1$
 \ba
 \label{CoulFF2}
 f_c(s,q^2)+f_c^{FF}(s,q^2) &=& -\frac{2\eta}{q^2}F_Q^2(q^2)e^{i \Phi_{c,FF}}, \\
 \label{PhicFF}
\Phi_{c,FF}(s,q^2) &\approx& -\eta \left(\frac{(q^2+\mu^2)^4}{\mu^8}-1\right)\ln{\frac{q^2}{q^2+\mu^2}}.
 \ea

 We can separate $f_{N,c}(s,q^2)$ which includes the effects of the nuclear scattering as modified by the Coulomb and form factor effects into two terms
\be
\label{CFFN}
f_{N,c}(s,q^2) = f_{N,c}^{\rm Corr}(s,q^2) + f_N(s,q^2) \, ,
\ee
where $f_N(s,q^2)$ is the pure nuclear amplitude and $ f_{N,c}^{\rm Corr}$ isolates the pieces of the full amplitude which involve both Coulomb-plus-form-factor and nuclear terms in a single small term. This is given for $pp$ scattering by
\be
\label{fppCorr}
 f_{N,c}^{\rm Corr}(s,q^2) = i \int_0^\infty db\,b \left( {\rm exp}  \left[ 2i\eta\gamma+i\eta\,{\rm ln}(\frac{q^2 b^2}{4})+2i\eta \sum_{m=0}^3 \frac{ (\mu b)^m}{2^m\Gamma(m+1)}K_m(\mu b) \right]-1 \right) \left(1-e^{2i\delta_N(b,s)}\right) J_0(qb) \, .
\ee
For $\bar{p}p$ scattering, we just change $\eta$ to $-\eta$ and use a relevant $\delta_N$ for the $\bar{p}p$ scattering.

For very small $q^2$ ($q^2 \leq 0.2$ GeV$^2$),  the real and imaginary parts of $ f_{N,c}^{\rm Corr}$ can be fitted using the following parametrization. For example, the real part  of $ f_{N,c}^{\rm Corr}$ is
 \be
 \label{fppCorr_fit}
\Re {f_{N,c}^{\rm Corr}}(s,t) = -( a_0+ b_0 \, {\rm ln}\,{p}) \,{\rm ln}\,t+( a_1+ b_1 {\rm ln}\,{p})+( a_2+ b_2  \, {\rm ln}\,{p} ) t + ( a_3+ b_3  \, {\rm ln}\,{p} ) t^2 \, ,
 \ee
where $a_i$ and $b_i$, $(i=0,1,2,3)$,  are the parameters whose units are  $\sqrt{\rm mb}/ {\rm GeV}$,  $\sqrt{\rm mb}/{\rm GeV}$,  $\sqrt{\rm mb}/{\rm GeV}^3$,  and $\sqrt{\rm mb}/{\rm GeV}^5$, for $i=0,1,2$ and $3$, respectively. The imagimary part will be fitted using the same form with a different set of the parameters.
 The parameters in the fit are given in Table \ref{table1:fppCorrfit}.

\begin{table}[ht]                   
%
\def\arraystretch{1.15}            
\begin{center}				  
\begin{tabular}[b]{|l|c|c|c|c|c|c|c|c|}

\hline
{\rm Parameters} & $a_0$ &  $b_0$  & $a_1$ &  $b_1$ & $a_2$ &  $b_2$& $a_3$ &  $b_3 $ \\
\hline
$pp$ Real part& $0.00553$ &  $0.01856$  & $0.01998$ &  $-0.07130$ & $-0.56648$ &  $0.33070$& $1.5591$ &  $-0.54019$ \\
 $pp$ Imaginary part& $0.00497$ &  $-0.00353$  & $-0.021147$ &  $0.01442$ & $0.12403$ &  $-0.07913$& $1.5591$ &  $0.15702$ \\
\hline
$\bar{p}p$  Real part& $-0.00743$ &  $-0.01842$  & $-0.01278$ &  $0.07083$ & $0.53333$ &  $-0.32971$& $-1.5049$ &  $0.54266$ \\
 $\bar{p}p$  Imaginary part & $-0.00471$ &  $0.00268$  & $0.02029$ &  $-0.01061$ & $-0.12343$ &  $0.05257$& $0.24633$ &  $-0.08269$ \\
\hline
\end{tabular}
     \caption{The parameters in the fit in \eq{fppCorr_fit} to the real and imaginary parts of  $ f_{N,c}^{\rm Corr}$.  \label{table1:fppCorrfit}}
\end{center}
\end{table}
\def\arraystretch{1}  
\noindent  We note that the results for the corrections are essentially model independent: any model that fits the data at small $q^2$, respects unitarity, and has the correct analytic phase structure will give the same correction to the accuracy needed.

We can now write the full amplitude in a form
\be
f(s,q^2) = f_1(s,q^2)+f_N(s,q^2),
\label{full_amp}
\ee
where the purely nuclear amplitude appears explicitly, with the Coulomb amplitude and the mixed Coulomb-nuclear corrections in $f_1$
\be
f_1(s,q^2) =  f_c(s,q^2)+f_c^{FF}(s,q^2) +  f_{N,c}^{\rm Corr}(s,q^2) \, . 
\label{f1_amp}
\ee

With our normalization, the differential elastic scattering amplitude is
\ba
\label{dsigma/dq^2}
\frac{d\sigma}{dq^2}(s,q^2)&=& \pi\lvert f(s,q^2)\rvert^2 \\
 &=& \pi \left(|f_1|^2+\frac{2|f_1||f_N|}{(1+\rho^2)^{1/2}}\, ({\rm sin} \Phi_1 + \rho \, {\rm cos} \Phi_1 ) + |f_N|^2 \right ) \, ,
\ea
where $\Phi_1$ is the phase of $f_1(s,q^2)$ and $\rho(s,q^2) = \Re f_N(s,q^2)/ \Im f_N(s,q^2)$. As mentioned in \cite{DH2020}, at high energies, $\Re f_N$ has a zero at small $q^2$, $\Re f_N(s,q_R^2)=0$ and, similarly, $\Im f_N(s,q_I^2)=0$ at the first diffraction dip in $d\sigma/dq^2$ at $q_I^2>q_R^2$. The zero in $\Re f_N$  is the expected Martin zero \cite{Martin}. The possible effects of this zero were discussed in Kohara {\it et al.} \cite{ Koharaetal} and included in the analysis of Pacetti {\it et al.} \cite{Pacettietal}, and were included in our recent work \cite{DH2020} as well. We showed in fact that  a reasonable approximate expression for the eikonal $\rho(s,q^2$) from $q^2=0$ through the region of the Martin zero was
$\rho(s,q^2) \approx \rho(s)\,\frac{1-q^2/q_R^2}{1-q^2/q_I^2}$  where $q_R$ and $q_I$  are the locations of the zeros in $\Re f_N$ and $\Im f_N$, respectively.


\section{Applications}
\subsection{Fits to the differential cross sections \label{sec:application}}


As an application of the above-mentioned results, we consider the model in \eq{curvature_exp} which has been used frequently in the analysis of experimental data, {\em e.g.}, the TOTEM data at 8 and 13 TeV and the ATLAS data at 13 TeV; see \cite{TOTEM2016,TOTEM2019,ATLAS2022} and earlier references therein. In this model, the phase of the nuclear amplitude is taken as a constant independent of $q^2$. It is determined simply by the ratio  $\rho$ of the  real to the imaginary parts of the nuclear amplitude in the forward direction, corresponding to a phase $\Phi_N(s,q^2)$ and $f_N(s,q^2)=e^{i\Phi_N}\lvert f_N(s,q^2)\rvert$.

From  \eq{curvature_exp}, taking the square root and introducing the phase of the nuclear amplitude $\Phi_N=\frac{\pi}{2}-\arctan{\rho(s,q^2)}$, we have \footnote{We have considered the variation of $\rho$ in the earlier paper  \cite{DH2020} taking into account the nearby diffraction zero in the real part of the amplitude, but that this $q^2$ dependence of $\rho$ does not significantly affect our results because  the interference effects which determine $\rho$ are limited to a very small range of $q^2$ near $q^2=0$.}
\be
\label{fNcurv}
\sqrt{\pi}f_N(s,q^2) \approx \sqrt{A}\,e^{i\Phi_N}e^{-\frac{1}{2}\left(Bq^2-Cq^4+Dq^6-\cdots\right)}.
\ee
%

\begin{figure}[htbp]
\includegraphics{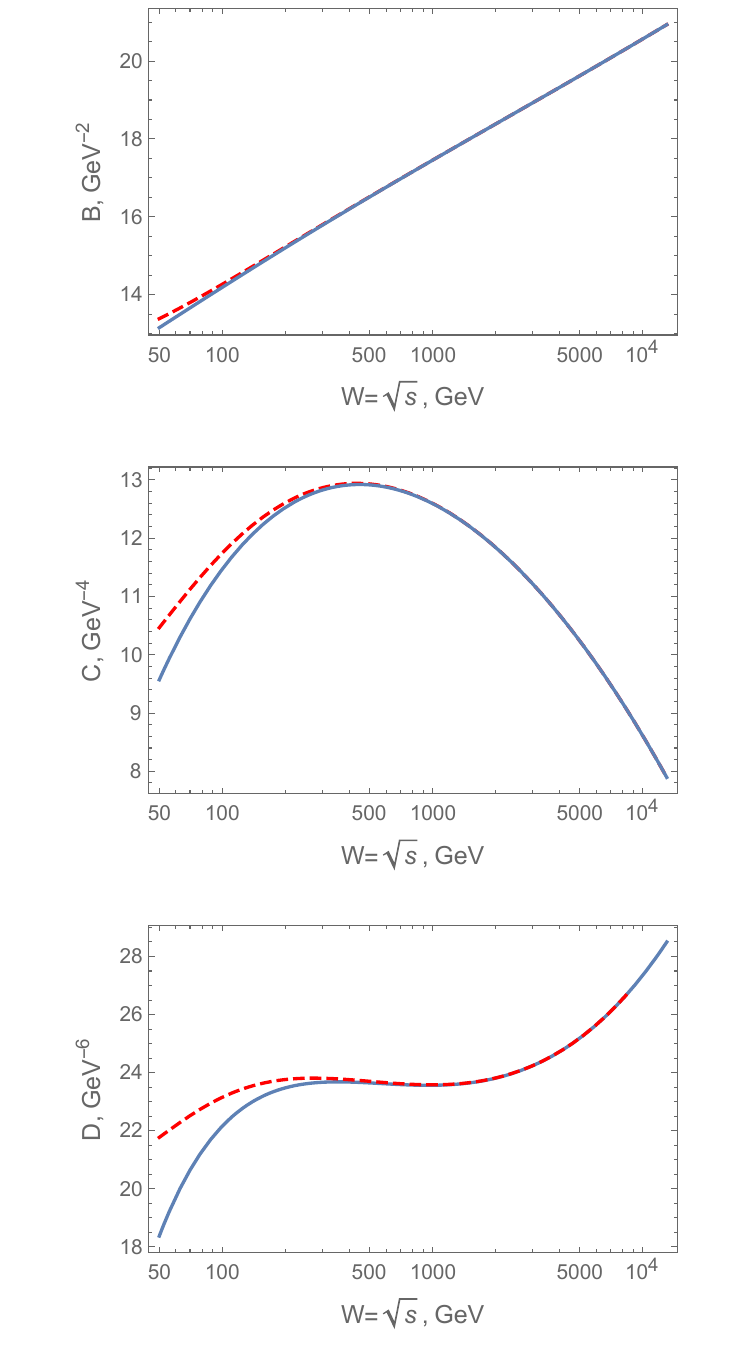}
\caption{Plot of the values of $B$, $C$, and $D$, calculated using our eikonal approach, versus $W=\sqrt{s}$ for the local momentum transfer $q_0^2=10^{-6}$ GeV$^2$  for $pp$ (solid blue curves) and $\bar{p}p$ (dashed red curves). The behavior at the lower energies is largely the result of the importance of the Regge-like terms in the eikonal function at lower energies.  For each parameter $B$, $C$, and $D$, respectively, the curves for $pp$ and $\bar{p}p$ are identical at high energies as expected.}
\label{fig:BCDvsW}
\end{figure}

Here we use the form of the hadronic cross section in \eq{curvature_exp}, with $C$ and $D$ taken from our eikonal results \footnote{The expansion in \eq{curvature_exp} and its range of validity were investigated in detail in \cite{bdhh-curvature}, where exact expressions were given for the parameters $B,\,C,\,{\rm and}\ D$ in the eikonal approach. As noted there, the predicted values of those parameters were consistent with the results obtained by TOTEM Collaboration in their fits to their TeV data \cite{TOTEM2015}.}  and $A$,$ B$, and $\rho$ used as the fitting parameters, to reanalyze the experimental data at 53 GeV, 62.3 GeV,  8 TeV, and 13 TeV. We have included the Coulomb scattering corrections and the Coulomb-hadronic interference terms in these fits. Our eikonal results give $C =9.770$ GeV$^{-4}$, $D=18.83$ GeV$^{-6}$ at 53 GeV, $C =10.29$ GeV$^{-4}$, $D=19.98$ GeV$^{-6}$ at 62.3 GeV,  $C =9.176$ GeV$^{-4}$,  $D=26.53$ GeV$^{-6}$ at 8 TeV, and  $C =7.896$ GeV$^{-4}$,  $D=28.50$ GeV$^{-6}$  at 13 TeV, respectively.  In \fig{fig:BCDvsW}, we show our calculated values of $B$, $C$, and $D$ as functions of the center-of-mass energy $W=\sqrt{s}$ for the local momentum transfer $q_0^2=10^{-6}$ GeV$^2$ for $pp$ (solid blue curves) and $\bar{p}p$ (dashed red curves). The curves are cut off at 50 GeV at the lower end. The behavior at the lower energies is largely the result of the importance of the Regge-like terms in the eikonal function at lower energies. For each parameter $B$, $C$, and $D$, respectively, the curves for $pp$ and $\bar{p}p$ are identical at high energies as expected.
%

We have analyzed the fits using data up to a maximum value $q^2_{\rm max}$.  We emphasize that our analyses are based on straightforward least squares fits to the rather precise data at those energies using only the quoted statistical errors.  In Table \ref{table:fitparameters}, we summarize the results of our fits to the ISR data at 53 GeV and 62.3 GeV for $q^2_{\rm max}=0.1$  GeV$^2$ and fits to TOTEM data at 8 TeV and 13 TeV for $q^2_{\rm max}=0.07$ GeV$^2$, $0.10$ GeV$^2$, and $0.15$ GeV$^2$. For each of the fits presented there, the total cross-section was derived via the optical theorem:
\be
\label{sigma_tot}
\sigma^2_{\rm tot} = {16 \pi A \over 1+ \rho^2} \, .
\ee
%


 \begin{table}[ht]                   
 \caption{\protect  The results of our fits to the ISR data at 53 GeV and 62.3 GeV  \cite{ Nagy1979, Breakstone1985, ISR1985} and to the TOTEM data at 8 TeV and 13 TeV  \cite{TOTEM2016,TOTEM2019}. The Coulomb and Coulomb-hadronic interference contributions to the scattering were included in the fit. $A$, $B$, and $\rho$ are the corresponding parameters in  fits which included the curvature parameters $C$ and $D$, with  $(d\sigma/dq^2)_N \approx A \exp(-Bq^2+Cq^4-Dq^6)$. The parameters $C$ and $D$  were calculated using the comprehensive eikonal fit to the high energy $pp$ and $\bar{p}p$ data in \cite{curvature_fit}.   \label{table:fitparameters}}				 
\def\arraystretch{1.15}            
\begin{center}	
\begin{ruledtabular}			  
\begin{tabular}[b]{|c|c|c|c|c|c|c||c|}
(GeV$^2$) & $W$ ({\rm GeV})& d.o.f & $\chi^2/{\rm d.o.f.}\ $  & $A$\ ( mb/GeV$^{2}$) & $B$\ ( GeV$^{-2}$)\   & $\rho$ & $\sigma_{\rm tot}($\rm mb$) $\ \\
 &  & & & & & & \\
\hline
 & & & & & & & \\
1) $q^2_{\rm max}=$0.07 &  13000 & 76   & $0.869\ $ & $647.2 \pm 0.7\ $ & $21.23\pm 0.03\ $ & $0.095 \pm 0.004\ $ & $112.0 \pm 0.1\ $ \\
&  8000 & 15  & $0.775\ $ & $552.3 \pm 2.9\ $ & $20.68\pm 0.12\ $  & $0.105 \pm 0.020\ $ &  $103.4 \pm 0.3\ $\\
\hline
 & & & & & & & \\
2) $q^2_{\rm max}=$0.10 & 13000 & 93   & $0.956\ $ & $645.8 \pm 0.6\ $ & $21.16\pm 0.02\ $ & $0.091 \pm 0.004\ $ &  $112.0 \pm 0.1\ $ \\
 & 8000 & 18  & $0.710\ $ & $551.5 \pm 2.5\ $ & $20.63\pm 0.08\ $  & $0.102 \pm 0.019\ $ &  $103.4 \pm 0.3\ $ \\
 & 62.3 & 19  & $1.448\ $ & $97.56 \pm 0.72\ $ & $13.40\pm 0.18\ $  & $0.071 \pm 0.018\ $ &  $43.59 \pm 0.17\ $ \\
 & 53 & 18  & $2.048\ $ & $92.98 \pm 0.21\ $ & $13.40\pm 0.07\ $  & $0.082 \pm 0.002\ $ &  $42.52 \pm 0.05\ $ \\
\hline
 & & & & & & & \\
3) $q^2_{\rm max}=$0.15 & 13000 & 116 & $1.290\ $ & $644.1 \pm 0.5\ $ & $21.09\pm 0.01\ $ & $0.085 \pm 0.004\ $ &  $111.9 \pm 0.1\ $ \\
&  8000 & 23  & $1.330\ $ & $547.3 \pm 2.1\ $ & $20.43\pm 0.06\ $  & $0.086 \pm 0.018\ $ & $103.1 \pm 0.3\ $ \\
\end{tabular}
\end{ruledtabular}

\end{center}
\end{table}
\def\arraystretch{1}
 %

\begin{figure}[htbp]
\includegraphics{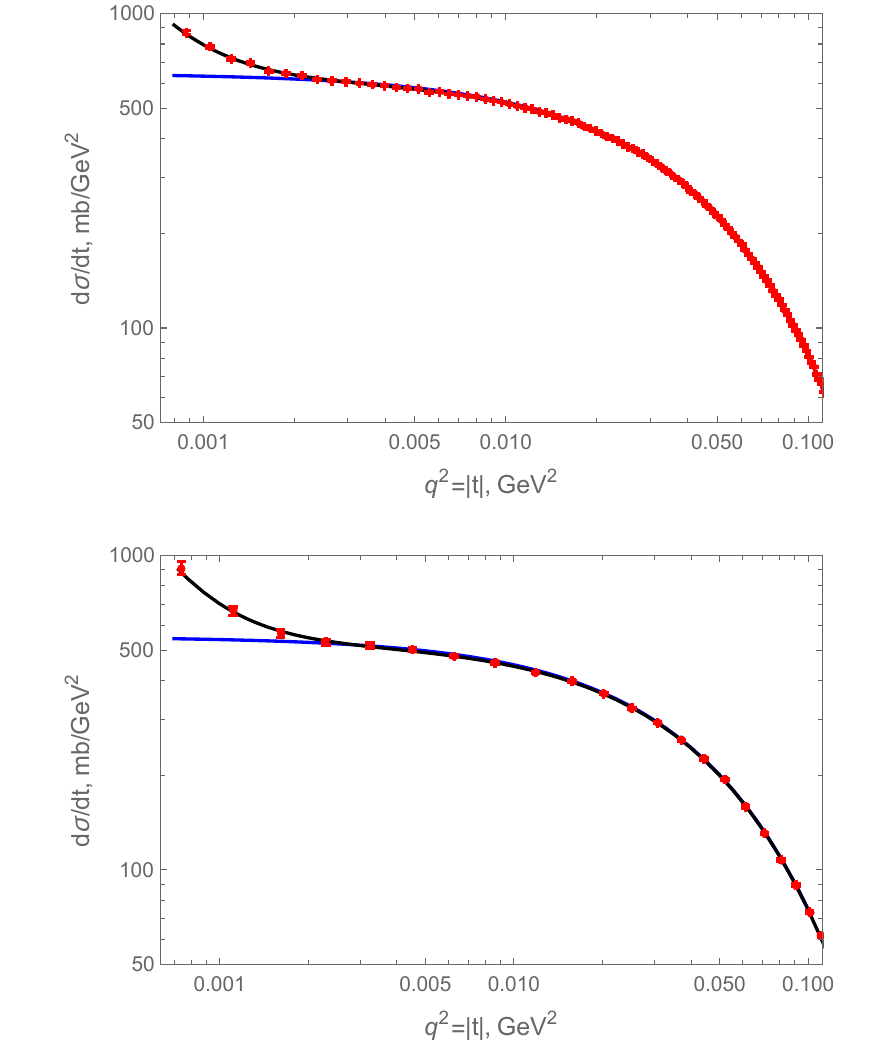}
\caption{Fits to the differential $pp$ elastic scattering cross sections $d\sigma/dq^2$ for the TOTEM data at 8 TeV (bottom panel) and 13 TeV (top panel) over the interval $q^2 \le 0.10$ GeV$^2$ . The values of the curvature terms $C$ and $D$ in the series expansion of the hadronic contribution to $\ln(d\sigma/dq^2)$ were taken from the overall eikonal fit to the high-energy $pp$ and $\bar{p}p$ data in \cite{curvature_fit}.  $d\sigma/dq^2$ from the fit and the purely nuclear result of the fit are given by the black and blue curves, respectively. Data with their statistical errors are red. }
\label{fig:TOTEM_dxsecs}
\end{figure}

For the TOTEM data at 8 TeV and 13 TeV,  the fits for case 1)  ($q^2_{\rm max}=$0.07) and case 2)  ($q^2_{\rm max}=$0.10) are all excellent with $\chi^2/{\rm d.o.f.}$ less than 1. The results from the fits are consistent with those found by the TOTEM Collaboration. As an illustration, we show the fits to $d\sigma/dq^2$ for the TOTEM data at 8 TeV and 13 TeV in \fig{fig:TOTEM_dxsecs} over the interval $q^2 \le 0.10$ GeV$^2$  in conventional logarithmic plots.


%

We also use  \eq{curvature_exp} to study the case of $C=D=0$  (i.e., no curvature corrections) for the TOTEM data at 8 TeV and 13 TeV. We find that omitting $D$ and $C$ does not change $\rho$ much because the only sensitivity to $\rho$ is at very small $q^2$ where $C$ and $D$ terms are very small.  However, the cross section fits over the region shown in  \fig{fig:TOTEM_dxsecs} would be much worse if they are omitted. 

Doing some fits with our variable $\rho(s,q^2)$ in the earlier paper \cite{DH2020} for $q^2<0.1$ GeV$^2$, we have found that the fit results and the fitting parameters do not change noticeably relative to the present results. The possible influence on $\rho$ of the Martin zero in $\Re f_N$  is one of the main points  in Kohara {\it et al.} \cite{ Koharaetal} and Pacetti {\it et al.} \cite{Pacettietal}. In particular, Pacetti {\it et al.} get a significant effect not by using their $q^2$ - dependent expression and fitting with that, but simply note that using an average of the variable result over the fitting range leads to a smaller fitted $\rho$.  Like us, others have found no significant change when $\rho(s,q^2)$ is allowed to vary. This is because the sensitivity to the Coulomb-nuclear interference in the fitting is only at very small $q^2$.

\subsection{ The differential cross section $d \sigma/dq^2$ in the dip region \label{sec:application2}}

In order to see what differences one would expect in the eikonal model with Coulomb corrections, we look at the region near the observed dip in the differential cross section $d \sigma/dq^2$ for $pp$ and $\bar{p}p$ elastic scattering. It is known that the cross sections in the dip region are very sensitive to small effects, and the data, especially the $pp$ and $\bar{p}p$ differences which are affected by different normalization uncertainties on the two cross sections, are uncertain there. Thus, one cannot really specify an experimental difference at a point.

\begin{figure}[htbp]
\includegraphics{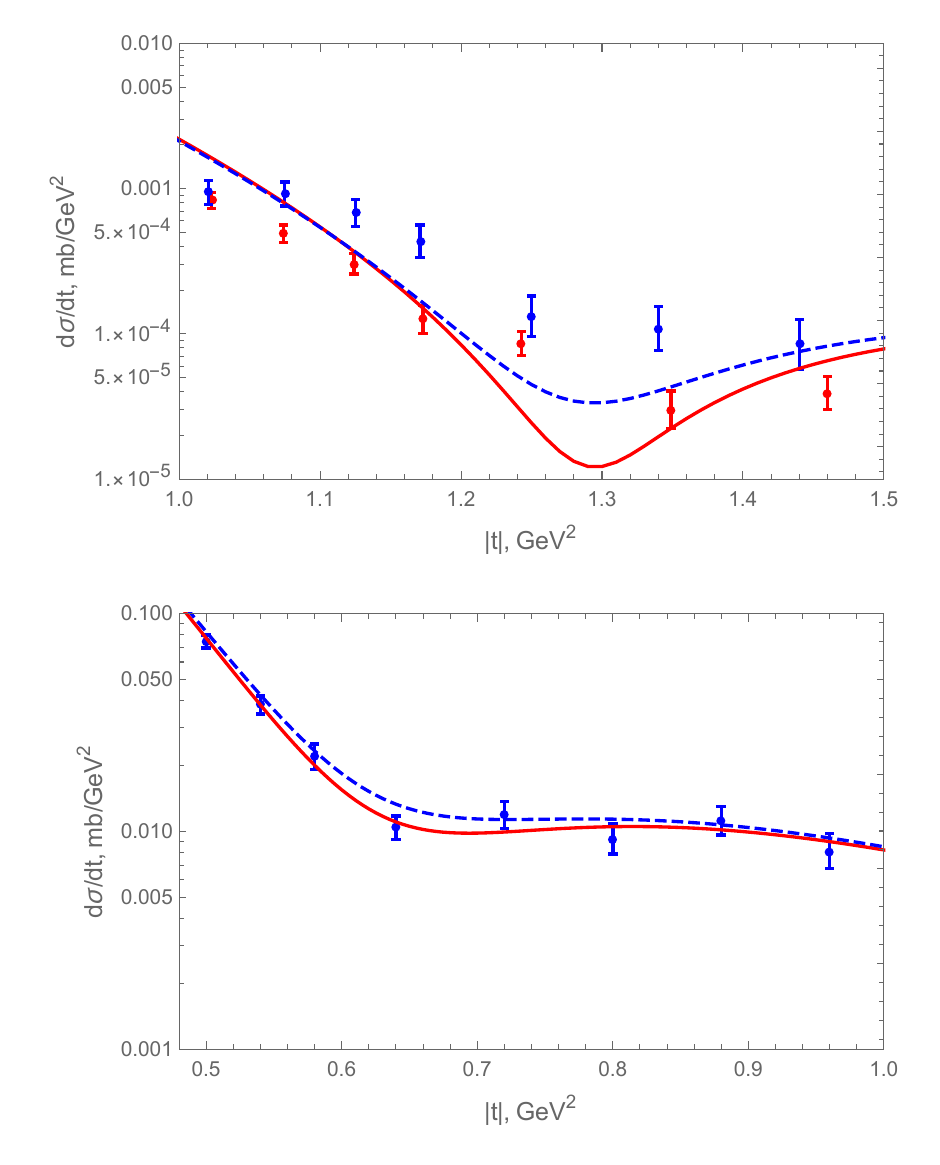}
\caption{Differential cross sections  $d \sigma/dq^2$ in the dip region for $pp$ and $\bar{p}p$ at  $W =$ 53 GeV (top panel) and 1.96 TeV (bottom panel).  The red, solid curve and the blue, dashed curve are the differential cross sections $d \sigma/dq^2$ for $pp$ and $\bar{p}p$ from our eikonal model with Coulomb corrections.   Note that data for $pp$ (red) and $\bar{p}p$ (blue) with their statistical errors at $W =$ 53 GeV  \cite{ Nagy1979, Breakstone1985} are shown in the top panel and that only data for $\bar{p}p$ (blue) with their statistical errors at $W =$ 1.96 TeV \cite{D02012} are available in the bottom panel. }
\label{fig:dsigmadt_dipregion}
\end{figure}

 In \fig{fig:dsigmadt_dipregion}, we plot the differential cross sections in the dip region for $pp$ and $\bar{p}p$ at $W =$ 53 GeV (top panel) and 1.96 TeV (bottom panel). In the figure, the theoretical differential cross sections $d \sigma/dq^2$ from our eikonal model with Coulomb corrections for $pp$ and $\bar{p}p$ are the red, solid curve and the blue, dashed curve, respectively.  Note that data for $pp$ (red) and $\bar{p}p$ (blue) with their statistical errors at $W =$ 53 GeV  \cite{ Nagy1979, Breakstone1985} are shown in the top panel and that only data for $\bar{p}p$ (blue) with their statistical errors at $W =$ 1.96 TeV  \cite{ D02012} are available in the bottom panel. 

We see that, at $W =$ 53 GeV,  our theoretical curves for $\bar{p}p$ and $pp$ show a dip from the first diffraction zero in the dominant imaginary part of the nuclear scattering amplitude. This is predicted to be at $|t|$ = 1.295 GeV$^2$ which is close to the observed minimum \cite{ Nagy1979}. However, the ratio $d \sigma^{\bar{p}p}/d \sigma^{pp}$=2.76  evaluated at the dip location  $|t|$ = 1.295 GeV$^2$ is smaller than the experimental value 4.5 $\pm$ 1.7 at $|t|$ = 1.333 GeV$^2$ quoted in \cite{ Breakstone1985}. While some very small adjustment of the nuclear amplitude would be needed to obtain a more precise fit to the data, that is unlikely to change the conclusion that Coulomb-nuclear interference effects are significant in this region where the amplitude becomes predominantly real, and that those effects lead to a very significant difference between the $pp$ and $\bar{p}p$  cross sections near the dip.

At $W =$ 1.96 TeV, both our theoretical curves for $pp$ and $\bar{p}p$ seem to agree with the D0 data for $\bar{p}p$ \cite{ D02012} but the curves do not show clearly a dip location. The predicted location of the zero in the imaginary part of the $pp$ amplitude is $|t|$ = 0.638 GeV$^2$.  There are currently no $pp$ data at 1.96 TeV, and the only information on cross section differences comes from the D0/TOTEM work using extrapolated $pp$ cross sections \cite{ D0TOTEM2020}.  We find that the Coulomb effects in the dip region at 1.96 TeV are still significant on the scale of the projected differences given in \cite{ D0TOTEM2020}, Fig. 4, and should be included in analyzing those differences.

We conclude that Coulomb effects are significant in the dip region at 53 GeV and 1.96 TeV, and must be taken into account in searches for odderon effects through cross section differences in that region and energy range. The interference effects in the dip region are smaller at higher energies, but are still significant. For example, the difference at 7 TeV is about 20\% of the mean cross section at the dip at $|t|$ = 0.478 GeV$^2$,  about the same as the statistical uncertainty in the measured $\bar{p}p$ cross section. 

%


\section{Conclusions \label{sec:conclusions}}

Using the simple eikonal approach for Coulomb-nuclear interference and form-factor effects and taking into account the curvature corrections in proton-proton scattering, we have fitted the parameters $B$, $\rho$, and $\sigma_{\rm tot}$ at $\sqrt s =$ 53 GeV, 62.3 GeV,  8 TeV, and 13 TeV. We find that the results of the basic parameters calculated using our simple eikonal approach agree well with the results obtained using other methods.

 We have also investigated the differential cross sections in the dip region for $pp$ and $\bar{p}p$  elastic scattering at $W=$ 53 GeV and 1.96 TeV and find that Coulomb effects are significant in the dip region at 53 GeV and 1.96 TeV, and must be taken into account in searches for odderon effects through cross section differences in that energy region.

\begin{acknowledgments}
The author would like to thank Professor Loyal Durand for useful comments and invaluable support.
\end{acknowledgments}

\bibliography{CoulombMathbibPH}

\begin{thebibliography}{29}%
\makeatletter
\providecommand \@ifxundefined [1]{%
 \@ifx{#1\undefined}
}%
\providecommand \@ifnum [1]{%
 \ifnum #1\expandafter \@firstoftwo
 \else \expandafter \@secondoftwo
 \fi
}%
\providecommand \@ifx [1]{%
 \ifx #1\expandafter \@firstoftwo
 \else \expandafter \@secondoftwo
 \fi
}%
\providecommand \natexlab [1]{#1}%
\providecommand \enquote  [1]{``#1''}%
\providecommand \bibnamefont  [1]{#1}%
\providecommand \bibfnamefont [1]{#1}%
\providecommand \citenamefont [1]{#1}%
\providecommand \href@noop [0]{\@secondoftwo}%
\providecommand \href [0]{\begingroup \@sanitize@url \@href}%
\providecommand \@href[1]{\@@startlink{#1}\@@href}%
\providecommand \@@href[1]{\endgroup#1\@@endlink}%
\providecommand \@sanitize@url [0]{\catcode `\\12\catcode `\$12\catcode
  `\&12\catcode `\#12\catcode `\^12\catcode `\_12\catcode `\%12\relax}%
\providecommand \@@startlink[1]{}%
\providecommand \@@endlink[0]{}%
\providecommand \url  [0]{\begingroup\@sanitize@url \@url }%
\providecommand \@url [1]{\endgroup\@href {#1}{\urlprefix }}%
\providecommand \urlprefix  [0]{URL }%
\providecommand \Eprint [0]{\href }%
\providecommand \doibase [0]{https://doi.org/}%
\providecommand \selectlanguage [0]{\@gobble}%
\providecommand \bibinfo  [0]{\@secondoftwo}%
\providecommand \bibfield  [0]{\@secondoftwo}%
\providecommand \translation [1]{[#1]}%
\providecommand \BibitemOpen [0]{}%
\providecommand \bibitemStop [0]{}%
\providecommand \bibitemNoStop [0]{.\EOS\space}%
\providecommand \EOS [0]{\spacefactor3000\relax}%
\providecommand \BibitemShut  [1]{\csname bibitem#1\endcsname}%
\let\auto@bib@innerbib\@empty
\bibitem [{\citenamefont {Durand}\ and\ \citenamefont {Ha}(2020)}]{DH2020}%
  \BibitemOpen
  \bibfield  {author} {\bibinfo {author} {\bibfnamefont {L.}~\bibnamefont
  {Durand}}\ and\ \bibinfo {author} {\bibfnamefont {P.}~\bibnamefont {Ha}},\
  }\bibfield  {title} {\bibinfo {title} {Coulomb-nuclear interference effects
  in proton-proton scattering: A simple new eikonal approach},\ }\href@noop {}
  {\bibfield  {journal} {\bibinfo  {journal} {Phys. Rev. D}\ }\textbf {\bibinfo
  {volume} {102}},\ \bibinfo {pages} {036025} (\bibinfo {year}
  {2020})}\BibitemShut {NoStop}%
\bibitem [{\citenamefont {Bethe}(1958)}]{BetheCoulomb}%
  \BibitemOpen
  \bibfield  {author} {\bibinfo {author} {\bibfnamefont {H.~A.}\ \bibnamefont
  {Bethe}},\ }\bibfield  {title} {\bibinfo {title} {Scattering and polarization
  of protons by nuclei},\ }\href@noop {} {\bibfield  {journal} {\bibinfo
  {journal} {Ann. Phys. (NY)}\ }\textbf {\bibinfo {volume} {3}},\ \bibinfo
  {pages} {190} (\bibinfo {year} {1958})}\BibitemShut {NoStop}%
\bibitem [{\citenamefont {West}\ and\ \citenamefont
  {Yennie}(1968)}]{WestYennie}%
  \BibitemOpen
  \bibfield  {author} {\bibinfo {author} {\bibfnamefont {G.~B.}\ \bibnamefont
  {West}}\ and\ \bibinfo {author} {\bibfnamefont {D.~R.}\ \bibnamefont
  {Yennie}},\ }\bibfield  {title} {\bibinfo {title} {Coulomb interference in
  high-energy scattering},\ }\href@noop {} {\bibfield  {journal} {\bibinfo
  {journal} {Physical Review}\ }\textbf {\bibinfo {volume} {172}},\ \bibinfo
  {pages} {1413} (\bibinfo {year} {1968})}\BibitemShut {NoStop}%
\bibitem [{\citenamefont {Cahn}(1982)}]{Cahn}%
  \BibitemOpen
  \bibfield  {author} {\bibinfo {author} {\bibfnamefont {R.}~\bibnamefont
  {Cahn}},\ }\bibfield  {title} {\bibinfo {title} {Coulombic-hadronic
  interference in an eikonal model},\ }\href@noop {} {\bibfield  {journal}
  {\bibinfo  {journal} {Z. Phys. C}\ }\textbf {\bibinfo {volume} {15}},\
  \bibinfo {pages} {253} (\bibinfo {year} {1982})}\BibitemShut {NoStop}%
\bibitem [{\citenamefont {Islam}(1964)}]{Islam}%
  \BibitemOpen
  \bibfield  {author} {\bibinfo {author} {\bibfnamefont {M.~M.}\ \bibnamefont
  {Islam}},\ }\bibfield  {title} {\bibinfo {title} {Bethe's formula for
  \uppercase{C}oulomb-nuclear interference},\ }\href@noop {} {\bibfield
  {journal} {\bibinfo  {journal} {Physical Review}\ }\textbf {\bibinfo {volume}
  {162}},\ \bibinfo {pages} {1426} (\bibinfo {year} {1964})}\BibitemShut
  {NoStop}%
\bibitem [{\citenamefont {Kundr\'{a}t}\ and\ \citenamefont
  {Lokaji\v{c}ek}(1994)}]{KL-Coulomb}%
  \BibitemOpen
  \bibfield  {author} {\bibinfo {author} {\bibfnamefont {V.}~\bibnamefont
  {Kundr\'{a}t}}\ and\ \bibinfo {author} {\bibfnamefont {M.}~\bibnamefont
  {Lokaji\v{c}ek}},\ }\bibfield  {title} {\bibinfo {title} {High-energy elastic
  scattering amplitude of unpolarized and charged hadrons},\ }\href@noop {}
  {\bibfield  {journal} {\bibinfo  {journal} {Z. Phys. C}\ }\textbf {\bibinfo
  {volume} {63}},\ \bibinfo {pages} {619} (\bibinfo {year} {1994})}\BibitemShut
  {NoStop}%
\bibitem [{\citenamefont {Buttimore}\ \emph {et~al.}(1978)\citenamefont
  {Buttimore}, \citenamefont {Gotsman},\ and\ \citenamefont
  {Leader}}]{Buttimore}%
  \BibitemOpen
  \bibfield  {author} {\bibinfo {author} {\bibfnamefont {N.~H.}\ \bibnamefont
  {Buttimore}}, \bibinfo {author} {\bibfnamefont {E.}~\bibnamefont {Gotsman}},\
  and\ \bibinfo {author} {\bibfnamefont {E.}~\bibnamefont {Leader}},\
  }\bibfield  {title} {\bibinfo {title} {Spin-dependent phenomena induced by
  electromagnetic-hadronic interference at high energies},\ }\href@noop {}
  {\bibfield  {journal} {\bibinfo  {journal} {Phys. Rev. D}\ }\textbf {\bibinfo
  {volume} {18}},\ \bibinfo {pages} {694} (\bibinfo {year} {1978})}\BibitemShut
  {NoStop}%
\bibitem [{\citenamefont {Kopeliovich}\ and\ \citenamefont
  {Tarasov}(2001)}]{KopeliovichTarasov}%
  \BibitemOpen
  \bibfield  {author} {\bibinfo {author} {\bibfnamefont {B.~Z.}\ \bibnamefont
  {Kopeliovich}}\ and\ \bibinfo {author} {\bibfnamefont {A.~V.}\ \bibnamefont
  {Tarasov}},\ }\bibfield  {title} {\bibinfo {title} {The \uppercase{C}oulomb
  phase revisited},\ }\href@noop {} {\bibfield  {journal} {\bibinfo  {journal}
  {Phys. Lett. B}\ }\textbf {\bibinfo {volume} {497}},\ \bibinfo {pages} {44}
  (\bibinfo {year} {2001})}\BibitemShut {NoStop}%
\bibitem [{\citenamefont {Block}(2006)}]{Blockrev}%
  \BibitemOpen
  \bibfield  {author} {\bibinfo {author} {\bibfnamefont {M.~M.}\ \bibnamefont
  {Block}},\ }\bibfield  {title} {\bibinfo {title} {Hadronic forward
  scattering: Predictions for the large hadron collider and cosmic rays},\
  }\href@noop {} {\bibfield  {journal} {\bibinfo  {journal} {Phys. Rep.}\
  }\textbf {\bibinfo {volume} {436}},\ \bibinfo {pages} {71} (\bibinfo {year}
  {2006})}\BibitemShut {NoStop}%
\bibitem [{\citenamefont {Block}\ and\ \citenamefont {Cahn}(1985)}]{BlockCahn}%
  \BibitemOpen
  \bibfield  {author} {\bibinfo {author} {\bibfnamefont {M.~M.}\ \bibnamefont
  {Block}}\ and\ \bibinfo {author} {\bibfnamefont {R.~N.}\ \bibnamefont
  {Cahn}},\ }\bibfield  {title} {\bibinfo {title} {High-energy $p\bar{p}$ and
  $pp$ forward elastic scattering and total cross sections},\ }\href@noop {}
  {\bibfield  {journal} {\bibinfo  {journal} {Rev. Mod. Phys.}\ }\textbf
  {\bibinfo {volume} {57}},\ \bibinfo {pages} {563} (\bibinfo {year}
  {1985})}\BibitemShut {NoStop}%
\bibitem [{\citenamefont {Pancheri}\ and\ \citenamefont
  {Srivastava}(2017)}]{PanSri}%
  \BibitemOpen
  \bibfield  {author} {\bibinfo {author} {\bibfnamefont {G.}~\bibnamefont
  {Pancheri}}\ and\ \bibinfo {author} {\bibfnamefont {Y.}~\bibnamefont
  {Srivastava}},\ }\bibfield  {title} {\bibinfo {title} {Introduction to the
  physics of the total cross section at \uppercase{LHC}},\ }\href@noop {}
  {\bibfield  {journal} {\bibinfo  {journal} {Eur. Phys. J. C}\ }\textbf
  {\bibinfo {volume} {77:150}} (\bibinfo {year} {2017})},\ \Eprint
  {https://arxiv.org/abs/arXiv:1610.1003 [hep-ph]} {arXiv:1610.1003 [hep-ph]}
  \BibitemShut {NoStop}%
\bibitem [{\citenamefont {Block}\ \emph {et~al.}(2015)\citenamefont {Block},
  \citenamefont {Durand}, \citenamefont {Ha},\ and\ \citenamefont
  {Halzen}}]{eikonal2015}%
  \BibitemOpen
  \bibfield  {author} {\bibinfo {author} {\bibfnamefont {M.~M.}\ \bibnamefont
  {Block}}, \bibinfo {author} {\bibfnamefont {L.}~\bibnamefont {Durand}},
  \bibinfo {author} {\bibfnamefont {P.}~\bibnamefont {Ha}},\ and\ \bibinfo
  {author} {\bibfnamefont {F.}~\bibnamefont {Halzen}},\ }\bibfield  {title}
  {\bibinfo {title} {Eikonal fit to $ pp$ and $\bar{p}p$ scattering and the
  edge in the scattering amplitude},\ }\href@noop {} {\bibfield  {journal}
  {\bibinfo  {journal} {Phys.\ Rev.\ D}\ }\textbf {\bibinfo {volume} {92}},\
  \bibinfo {pages} {014030} (\bibinfo {year} {2015})},\ \Eprint
  {https://arxiv.org/abs/arXiv:1505.04842v1 [hep-ph]} {arXiv:1505.04842v1
  [hep-ph]} \BibitemShut {NoStop}%
\bibitem [{\citenamefont {Nagy}\ \emph {et~al.}(1979)\citenamefont {Nagy} \emph
  {et~al.}}]{Nagy1979}%
  \BibitemOpen
  \bibfield  {author} {\bibinfo {author} {\bibfnamefont {E.}~\bibnamefont
  {Nagy}} \emph {et~al.},\ }\bibfield  {title} {\bibinfo {title} {Measurement
  of elastic proton-proton scattering at large momentum transfer at the
  \uppercase{CERN} intersecting storage rings},\ }\href@noop {} {\bibfield
  {journal} {\bibinfo  {journal} {Nucl. Phys. B}\ }\textbf {\bibinfo {volume}
  {150}},\ \bibinfo {pages} {221} (\bibinfo {year} {1979})}\BibitemShut
  {NoStop}%
\bibitem [{\citenamefont {Breakstone}\ \emph {et~al.}(1985)\citenamefont
  {Breakstone} \emph {et~al.}}]{Breakstone1985}%
  \BibitemOpen
  \bibfield  {author} {\bibinfo {author} {\bibfnamefont {A.}~\bibnamefont
  {Breakstone}} \emph {et~al.},\ }\bibfield  {title} {\bibinfo {title}
  {Measurement of $\bar{p}p$ and $pp$ elastic scattering in the dip region at
  $\sqrt s=$ 53 \uppercase{G}e\uppercase{V}},\ }\href@noop {} {\bibfield
  {journal} {\bibinfo  {journal} {Phys. Rev. Lett.}\ }\textbf {\bibinfo
  {volume} {54}},\ \bibinfo {pages} {2180} (\bibinfo {year}
  {1985})}\BibitemShut {NoStop}%
\bibitem [{\citenamefont {Amos}\ \emph {et~al.}(1985)\citenamefont {Amos} \emph
  {et~al.}}]{ISR1985}%
  \BibitemOpen
  \bibfield  {author} {\bibinfo {author} {\bibfnamefont {N.}~\bibnamefont
  {Amos}} \emph {et~al.},\ }\bibfield  {title} {\bibinfo {title} {Measurement
  of small angle antiproton-proton and proton proton elastic scattering at the
  \uppercase{CERN} intersecting storage rings},\ }\href@noop {} {\bibfield
  {journal} {\bibinfo  {journal} {Nucl. Phys. B}\ }\textbf {\bibinfo {volume}
  {262}},\ \bibinfo {pages} {689} (\bibinfo {year} {1985})}\BibitemShut
  {NoStop}%
\bibitem [{\citenamefont {Antchev}\ \emph {et~al.}(2016)\citenamefont {Antchev}
  \emph {et~al.}}]{TOTEM2016}%
  \BibitemOpen
  \bibfield  {author} {\bibinfo {author} {\bibfnamefont {G.}~\bibnamefont
  {Antchev}} \emph {et~al.} (\bibinfo {collaboration} {TOTEM Collaboration}),\
  }\bibfield  {title} {\bibinfo {title} {Measurement of elastic pp scattering
  at $\sqrt{s}$ = 8 \uppercase{T}e\uppercase{V} in the
  \uppercase{C}oulomb-nuclear interference region -- determination of the
  $\rho$ parameter and total cross section},\ }\href@noop {} {\bibfield
  {journal} {\bibinfo  {journal} {Eur. Phys. J. C}\ }\textbf {\bibinfo {volume}
  {76}},\ \bibinfo {pages} {661} (\bibinfo {year} {2016})},\ \Eprint
  {https://arxiv.org/abs/arXiv:1610.00603v1} {arXiv:1610.00603v1} \BibitemShut
  {NoStop}%
\bibitem [{\citenamefont {Antchev}\ \emph {et~al.}(2019)\citenamefont {Antchev}
  \emph {et~al.}}]{TOTEM2019}%
  \BibitemOpen
  \bibfield  {author} {\bibinfo {author} {\bibfnamefont {G.}~\bibnamefont
  {Antchev}} \emph {et~al.} (\bibinfo {collaboration} {TOTEM Collaboration}),\
  }\bibfield  {title} {\bibinfo {title} {First determination of the $\rho$
  parameter at $\sqrt{s}$ = 13 \uppercase{T}e\uppercase{V}},\ }\href@noop {}
  {\bibfield  {journal} {\bibinfo  {journal} {Eur. Phys. J. C}\ }\textbf
  {\bibinfo {volume} {79}},\ \bibinfo {pages} {785} (\bibinfo {year} {2019})},\
  \Eprint {https://arxiv.org/abs/arXiv:1812.04732 [hep-ex]} {arXiv:1812.04732
  [hep-ex]} \BibitemShut {NoStop}%
\bibitem [{\citenamefont {Durand}\ and\ \citenamefont
  {Ha}(2019)}]{curvature_fit}%
  \BibitemOpen
  \bibfield  {author} {\bibinfo {author} {\bibfnamefont {L.}~\bibnamefont
  {Durand}}\ and\ \bibinfo {author} {\bibfnamefont {P.}~\bibnamefont {Ha}},\
  }\bibfield  {title} {\bibinfo {title} {Eikonal and asymptotic fits to
  high-energy data for $\sigma$, $\rho$, and \uppercase{B}: An update with
  curvature corrections},\ }\href@noop {} {\bibfield  {journal} {\bibinfo
  {journal} {Phys.\ Rev.\ D}\ }\textbf {\bibinfo {volume} {99}},\ \bibinfo
  {pages} {014009} (\bibinfo {year} {2019})},\ \Eprint
  {https://arxiv.org/abs/arXiv:1810.11325 [hep-ph]} {arXiv:1810.11325 [hep-ph]}
  \BibitemShut {NoStop}%
\bibitem [{\citenamefont {Block}\ \emph {et~al.}(2016)\citenamefont {Block},
  \citenamefont {Durand}, \citenamefont {Ha},\ and\ \citenamefont
  {Halzen}}]{bdhh-curvature}%
  \BibitemOpen
  \bibfield  {author} {\bibinfo {author} {\bibfnamefont {M.~M.}\ \bibnamefont
  {Block}}, \bibinfo {author} {\bibfnamefont {L.}~\bibnamefont {Durand}},
  \bibinfo {author} {\bibfnamefont {P.}~\bibnamefont {Ha}},\ and\ \bibinfo
  {author} {\bibfnamefont {F.}~\bibnamefont {Halzen}},\ }\bibfield  {title}
  {\bibinfo {title} {Slope, curvature, and higher parameters in $pp$ and
  $\bar{p}p$ scattering, and the extrapolation of measurements of
  $d\sigma(s,t)/dt$ to $t=0$},\ }\href@noop {} {\bibfield  {journal} {\bibinfo
  {journal} {Phys. Rev. D}\ }\textbf {\bibinfo {volume} {93}},\ \bibinfo
  {pages} {114009} (\bibinfo {year} {2016})},\ \Eprint
  {https://arxiv.org/abs/arXiv:1605.00152 [hep-ph]} {arXiv:1605.00152 [hep-ph]}
  \BibitemShut {NoStop}%
\bibitem [{\citenamefont {Stenzel}(2022)}]{ATLAS2022}%
  \BibitemOpen
  \bibfield  {author} {\bibinfo {author} {\bibfnamefont {H.}~\bibnamefont
  {Stenzel}} (\bibinfo {collaboration} {ATLAS Collaboration}),\ }\bibfield
  {title} {\bibinfo {title} {Determination of the total cross section and
  $\rho$-parameter from elastic scattering in $pp$ collisions at $\sqrt{s}$=13
  \uppercase{T}e\uppercase{V} with the \uppercase{ATLAS} detector},\
  }\href@noop {} {\bibfield  {journal} {\bibinfo  {journal} {PoS ICHEP2022}\ ,\
  \bibinfo {pages} {803}} (\bibinfo {year} {2022})},\ \Eprint
  {https://arxiv.org/abs/arXiv:2209.11487 [hep-ex]} {arXiv:2209.11487 [hep-ex]}
  \BibitemShut {NoStop}%
\bibitem [{\citenamefont {Abazov}\ \emph {et~al.}(2021)\citenamefont {Abazov}
  \emph {et~al.}}]{D0TOTEM2020}%
  \BibitemOpen
  \bibfield  {author} {\bibinfo {author} {\bibfnamefont {V.}~\bibnamefont
  {Abazov}} \emph {et~al.} (\bibinfo {collaboration} {The D0 and TOTEM
  Collaborations}),\ }\bibfield  {title} {\bibinfo {title} {Odderon exchange
  from elastic scattering differences between $\bar{p}p$ and $pp$ data at 1.96
  \uppercase{T}e\uppercase{V} and from pp forward scattering measurements},\
  }\href@noop {} {\bibfield  {journal} {\bibinfo  {journal} {Phys. Rev. Lett.}\
  }\textbf {\bibinfo {volume} {127}},\ \bibinfo {pages} {062003} (\bibinfo
  {year} {2021})},\ \Eprint {https://arxiv.org/abs/arXiv:2012.03981 [hep-ex]}
  {arXiv:2012.03981 [hep-ex]} \BibitemShut {NoStop}%
\bibitem [{Note1()}]{Note1}%
  \BibitemOpen
  \bibinfo {note} {Magnetic-moment scattering does not contribute to the
  spin-independent part of the scattering amplitude. Its contribution to the
  spin-dependent amplitude is also suppressed by an angular factor proportional
  to $\protect \sqrt {q^2}$ at small $q^2$. The factor $1/q^2$ from the photon
  propagator therefore partially cancels and the amplitude is suppressed
  relative to the charge-scattering term for $q^2\rightarrow 0$. The magnetic
  form factors $F_M^2(q^2)$ further suppress the scattering at large $q$. As a
  result, the magnetic terms do not contribute significantly to the scattering
  in the region of interest here. See a detail discussion in \cite
  {DH2020}}\BibitemShut {NoStop}%
\bibitem [{\citenamefont {Martin}(1997)}]{Martin}%
  \BibitemOpen
  \bibfield  {author} {\bibinfo {author} {\bibfnamefont {A.}~\bibnamefont
  {Martin}},\ }\bibfield  {title} {\bibinfo {title} {A theorem on the real part
  of the high-energy scattering amplitude near the forward direction},\
  }\href@noop {} {\bibfield  {journal} {\bibinfo  {journal} {Phys. Lett. B}\
  }\textbf {\bibinfo {volume} {404}},\ \bibinfo {pages} {137} (\bibinfo {year}
  {1997})}\BibitemShut {NoStop}%
\bibitem [{\citenamefont {Kohara}\ \emph {et~al.}(2017)\citenamefont {Kohara},
  \citenamefont {Ferreira}, \citenamefont {Kodama},\ and\ \citenamefont
  {Rangel}}]{Koharaetal}%
  \BibitemOpen
  \bibfield  {author} {\bibinfo {author} {\bibfnamefont {A.~K.}\ \bibnamefont
  {Kohara}}, \bibinfo {author} {\bibfnamefont {E.}~\bibnamefont {Ferreira}},
  \bibinfo {author} {\bibfnamefont {T.}~\bibnamefont {Kodama}},\ and\ \bibinfo
  {author} {\bibfnamefont {M.}~\bibnamefont {Rangel}},\ }\bibfield  {title}
  {\bibinfo {title} {Elastic amplitudes studied with the \uppercase{LHC}
  measurements at 7 and 8 \uppercase{T}e\uppercase{V}},\ }\href@noop {}
  {\bibfield  {journal} {\bibinfo  {journal} {Eur. Phys. J. C}\ }\textbf
  {\bibinfo {volume} {77}},\ \bibinfo {pages} {877} (\bibinfo {year} {2017})},\
  \Eprint {https://arxiv.org/abs/arXiv:1709.05713 [hep-ph]} {arXiv:1709.05713
  [hep-ph]} \BibitemShut {NoStop}%
\bibitem [{\citenamefont {Pacetti}\ \emph {et~al.}(2019)\citenamefont
  {Pacetti}, \citenamefont {Srivastava},\ and\ \citenamefont
  {Pancheri}}]{Pacettietal}%
  \BibitemOpen
  \bibfield  {author} {\bibinfo {author} {\bibfnamefont {S.}~\bibnamefont
  {Pacetti}}, \bibinfo {author} {\bibfnamefont {Y.}~\bibnamefont
  {Srivastava}},\ and\ \bibinfo {author} {\bibfnamefont {G.}~\bibnamefont
  {Pancheri}},\ }\bibfield  {title} {\bibinfo {title} {Analysis and
  implications of precision near-forward \uppercase{TOTEM} data},\ }\href@noop
  {} {\bibfield  {journal} {\bibinfo  {journal} {Phys. Rev. D}\ }\textbf
  {\bibinfo {volume} {99}},\ \bibinfo {pages} {034014} (\bibinfo {year}
  {2019})},\ \Eprint {https://arxiv.org/abs/arXiv:1811.00499v2 [hep-ph]}
  {arXiv:1811.00499v2 [hep-ph]} \BibitemShut {NoStop}%
\bibitem [{Note2()}]{Note2}%
  \BibitemOpen
  \bibinfo {note} {We have considered the variation of $\rho $ in the earlier
  paper \cite {DH2020} taking into account the nearby diffraction zero in the
  real part of the amplitude, but that this $q^2$ dependence of $\rho $ does
  not significantly affect our results because the interference effects which
  determine $\rho $ are limited to a very small range of $q^2$ near
  $q^2=0$.}\BibitemShut {Stop}%
\bibitem [{Note3()}]{Note3}%
  \BibitemOpen
  \bibinfo {note} {The expansion in Eq.\protect \,(\ref {curvature_exp}) and
  its range of validity were investigated in detail in \cite {bdhh-curvature},
  where exact expressions were given for the parameters $B,\protect
  \,C,\protect \,{\protect \rm and}\ D$ in the eikonal approach. As noted
  there, the predicted values of those parameters were consistent with the
  results obtained by TOTEM Collaboration in their fits to their TeV data \cite
  {TOTEM2015}.}\BibitemShut {Stop}%
\bibitem [{\citenamefont {Abazov}\ \emph {et~al.}(2012)\citenamefont {Abazov}
  \emph {et~al.}}]{D02012}%
  \BibitemOpen
  \bibfield  {author} {\bibinfo {author} {\bibfnamefont {V.}~\bibnamefont
  {Abazov}} \emph {et~al.} (\bibinfo {collaboration} {D0 Collaboration}),\
  }\bibfield  {title} {\bibinfo {title} {Measurement of the differential cross
  section $d\sigma/dt$ in elastic $p\bar{p}$ scattering at $\sqrt s=$ 1.96
  \uppercase{T}e\uppercase{V}},\ }\href@noop {} {\bibfield  {journal} {\bibinfo
   {journal} {Phys. Rev. D}\ }\textbf {\bibinfo {volume} {86}},\ \bibinfo
  {pages} {012009} (\bibinfo {year} {2012})}\BibitemShut {NoStop}%
\bibitem [{\citenamefont {Antchev}\ \emph {et~al.}(2015)\citenamefont {Antchev}
  \emph {et~al.}}]{TOTEM2015}%
  \BibitemOpen
  \bibfield  {author} {\bibinfo {author} {\bibfnamefont {G.}~\bibnamefont
  {Antchev}} \emph {et~al.} (\bibinfo {collaboration} {TOTEM Collaboration}),\
  }\bibfield  {title} {\bibinfo {title} {Evidence for non-exponential elastic
  proton-proton differential cross section at low $\lvert t\rvert$ and
  $\sqrt{s}=8$ \uppercase{T}e\uppercase{V}},\ }\href@noop {} {\bibfield
  {journal} {\bibinfo  {journal} {Nucl. Phys. B}\ }\textbf {\bibinfo {volume}
  {899}},\ \bibinfo {pages} {527} (\bibinfo {year} {2015})}\BibitemShut
  {NoStop}%
\end{thebibliography}%

\end{document}